\documentclass[twocolumn,showpacs,preprintnumbers,amssymb]{revtex4}
\usepackage{graphicx}
\usepackage{dcolumn}
\usepackage{bm}

\begin{document}
\input epsf

\title{Richtmyer-Meshkov instability and solid $^4$He melting driven by acoustic pulse}
\author{N. Gov}
\address{Department of Materials and Interfaces,
The Weizmann Institute of Science,\\
P.O.B. 26, Rehovot, Israel 76100}

\begin{abstract}
Recent experiments have shown remarkable dynamics of solid $^4$He
melting and growth, driven by the normal incidence of an acoustic
pulse on the solid-liquid interface. The theory of solid
growth/melting, driven by the radiation pressure of the acoustic
pulse, accounts well for the temperature dependence of the
measured data. There is however an observed source of extra,
temperature-independent, melting. We here propose that this extra
melting is due to solid-liquid mixing (and consequent melting) at
the interface, in a process similar to the Richtmyer-Meshkov
instability: Initial undulations of the rough interface, grow when
accelerated by the acoustic pressure oscillations. This model
predicts a temperature-independent extra melting and its
dependence on the acoustic power, which is in agreement with the
measured data.
\end{abstract}

\pacs{67.80.-s,68.08.-p,43.35.+d,47.20.–k}

\maketitle

Recent experiments \cite{nomura} have shown that the radiation
pressure $P_r$ of an acoustic pulse, incident on the
solid-superfluid interface of $^4$He, can induce the local melting
or growth of the solid. The general temperature dependence of the
pulse-induced melting/growth is accounted for using the concept of
acoustic radiation pressure. Nevertheless, a large deviation was
found, which seems to be temperature independent. It is this extra
melting which we shall describe in this paper.

We begin with the analysis of the experimental data in terms of a
linear growth/melting coefficient (Eq.1-6 of \cite{nomura}). The
amplitude of melting/growth is linearly related to the radiation
pressure $P_r$ through the (temperature dependent) growth
coefficient $K(T)$ \cite{nomura}
\begin{equation}
h=K(T)\frac{P_r}{\rho_{c}}\tau
\label{amppres}
\end{equation}
where $\tau=1$msec is the overall duration of the acoustic pulse.
 The radiation pressure is itself  temperature dependent:
$P_{r}=E\left[
1-\frac{c_1}{c_2}+R^2\left(1+\frac{c_1}{c_2}\right)\right]$, where
the energy density of the incident wave is $E=I/c_1$, the power
density $I=144$W/m$^2$, and the reflection coefficient is
\begin{equation}
R=\frac{z_2-z_1-z_{1}z_{2}\rho_{1}K(T)\left(\frac{\rho_{1}-\rho_{2}}{\rho_{1}\rho_{2}}\right)^2}
{z_1+z_2+z_{1}z_{2}\rho_{1}K(T)\left(\frac{\rho_{1}-\rho_{2}}{\rho_{1}\rho_{2}}\right)^2}
\end{equation}
where $z_{1,2}=\rho_{1,2}c_{1,2}$. The growth coefficient $K(T)$
is given by
\begin{equation}
K(T)^{-1}=A T^4 + B e^{(\frac{-\Delta}{T})} \label{growthcoef}
\end{equation}
(with $A=3\times 10^{-2}$,$B=2.4\times 10^{3}$,$\Delta=7.2$K
\cite{bowley}). In these expressions $\rho_{1,2},c_{1,2}$ are the
density and sound velocities of the two phases on either side of
the interface, with the index 1 corresponding to the phase on the
side of the incident wave. In the experiment \cite{nomura} we have
$\rho_{l}=0.17$gr/cc, $c_{l}=365$m/sec and $\rho_{s}=0.19$gr/cc,
$c_{s}=475$m/sec. In Fig.1 we compare the experimental data with
the calculated growth/melting amplitude $h$, given by
Eq.(\ref{amppres}). The overall agreement is good (dotted lines in
Fig.1), but there is clearly an additional source of melting which
appears for pulses incident from both the solid and liquid sides.
This extra melting is not caused by the sound heating of the solid
\cite{nomura0}.

The contribution from the acoustic pressure leads to
growth/melting that proceeds at a constant rate for the duration
of the applied pulse (\ref{amppres}). In addition there is the
inertial acceleration of the solid-liquid interface by the
incident pressure oscillations. This motion should average to zero
if no instability or mixing occurs at the interface. It is only
due to mixing at the interface that this mechanism can contribute
on average to the position of the interface at the end of the
pulse. We assume here that the shape changes of the interface due
to Richtmyer-Meshkov (RM)-mixing \cite{rychtmyer} do not change
the average acoustic pressure and therefore can be approximately
decoupled from each other. We therefore add the contribution from
the acoustic pressure to the mixing contribution due to the
RM-mixing. A complete description has to take into account
possible interference effects of one mechanism with the other.

Let us now summarize the main results of this paper. We propose
that the measured extra melting is the result of dynamic
solid-liquid mixing at the interface. This mixing is due to a
RM-instability \cite{rychtmyer}, that typically occurs when two
liquids/gases of different densities are accelerated one against
the other by an incident shock-wave, causing initial perturbations
of the interface to grow. In our case the interface is between two
phases of the same material, but of different density, which is
repeatedly accelerated to and fro by a series of rapid pressure
oscillations. Under these conditions, and a rough initial
interface, solid-liquid mixing should occur due to a RM-type
instability \cite{rychtmyer}, since we note that in the limit of
weak shock-waves these behave as acoustic pulses \cite{landau}.
Using the linear growth theory of the RM instability we predict a
temperature-independent melting, with the correct dependence on
the incident acoustic power. With a single free parameter we get
good quantitative account of the measured extra melting.

A RM instability will induce a layer of mixing at the interface
between two materials of different densities (here the
solid-liquid interface). The mixing occurs when initial
perturbations of the interface between two liquids (or gases) grow
after the interface is accelerated by a shock-wave
\cite{rychtmyerjpg} (Fig.2). The linear single-mode theory
describes the inertial growth of an initial interface undulation,
due to an impacting shock-wave, gives \cite{rychtmyer}
\begin{equation}
\frac{da}{dt}=k a_{0} v A \Rightarrow a_{RM}=a_{0}+k a_{0} v A
\tau \label{growthrm}
\end{equation}
where $k$ is the wavevector of the interface undulation of initial
amplitude $a_{0}$, $v$ is the interface velocity due to the
acceleration and $A=(\rho_{2}-\rho_{1})/(\rho_{2}+\rho_{1})$ is
the Atwood number. The sign of $A$ relates to the phase-reversal
process (Fig.2b) and is irrelevant here. Note that
Eq.(\ref{growthrm}) is temperature-independent. Since in our case
the solid-liquid interface is continuously accelerated to and fro
by the pressure oscillations \cite{noziers2}, we therefore
interpret the equation for the rate of growth of surface
instabilities (\ref{growthrm}), as giving the rate at which a
continuous flux of solid is locally mixed with the liquid and
melts.

Furthermore, unlike the usual RM scenario, we have a solid crystal
on one side of the interface. Due to the rapid
melting-solidification at the rough (i.e. unfaceted)
solid-superfluid interface, this interface deforms similar to that
between two liquids \cite{andreev}. On the other hand, dissipation
and surface tension tend to damp out any inertial growth of
instabilities. The solid-liquid surface tension, is
$\sigma\sim0.2$erg/cm$^2$ for $^4$He \cite{surftenshe}, and acts
to stabilize and damp the inertial growth of initial perturbations
of the RM-type \cite{surftens}. The regime where the surface
tension dominates occurs whenever the damping rate
$\omega=k^{3/2}\sqrt{\sigma/(\rho_{1}+\rho_{2})}$ is larger than
the growth rate given in Eq.(\ref{growthrm}), i.e. for $k>(a_{0} v
A)^{2}(\rho_{1}+\rho_{2})/\sigma$. Using the relevant values
(estimating $a_{0}\lesssim 10$\AA), we find that in the
experiments the surface tension dominates. This means that
following a single pressure oscillation, the amplitude growth
described in (\ref{growthrm}) would have been quickly damped out,
and result in negligible solid-liquid mixing.

Nevertheless, in the experiment \cite{nomura}, there is not a
single pressure oscillation followed by inertial growth, but
rather a long pulse of continuous acoustic waves. Each pressure
oscillation (of $\sim0.1\mu$sec duration, frequency $9.9$MHz
\cite{nomura}) will cause the solid-liquid interface to
accelerate, and existing undulations to grow in amplitude. The
solid-liquid interface is therefore continuously accelerated back
and forth, throughout the pulse duration. The damping due to the
surface tension is therefore not relevant, and we proceed to
estimate the amount of solid-liquid mixing due to this
continuously driven RM-instability, using the simplest linear
growth theory \cite{rychtmyer} (Eq. \ref{growthrm}), over the
entire pulse duration $\tau$. In this case we treat the growth
rate equation (\ref{growthrm}) as a measure of the rate at which
solid material is continuously mixed into the liquid, in a manner
similar to the (highly exaggerated) "spike" in Fig.2. This
material then melts as it finds itself in a region which is liquid
in equilibrium. The accumulated amplitude $a_{RM}$ represents
therefore the total thickness of solid that was mixed (and melted)
by this process over the duration of the acoustic pulse. In
particular, it {\em does not} correspond to the final amplitude of
individual interface undulations.

The interface velocity $v$, due to the oscillating pressure of
amplitude $\delta P= \sqrt{E \rho_{1,2} c_{1,2}^2}$ (depending on
the direction of the incident wave), is given in terms of the
local material velocities at the interface, after the passing of
the pressure wave \cite{noziers2}:
$v=(\rho_{1}v_{m,1}-\rho_{2}v_{m,2})/(\rho_{1}-\rho_{2})$. The
local material velocities following a pressure oscillation are
given by \cite{noziers2} $v_{m}=\delta P /z_{1,2}$. For power
density $I=144$W/m$^2$ \cite{nomura} we therefore get: $v\simeq
0.023$m/sec.

In Fig.1 we compare the calculated combined melting/growth
$h-a_{RM}$ (\ref{amppres},\ref{growthrm}) with the experimental
data. We find that a good agreement \cite{note} is achieved when
we use an interface roughness aspect-ratio of $k a_{0}\sim 20$ in
Eq.(\ref{growthrm}). For a rough solid-liquid interface, which is
the case in the experiments done so far \cite{nomura,nomura0}, we
would naively expect $k a_{0}\sim 1$. This discrepancy indicates
that the linear growth theory is not entirely adequate, especially
as it usually deals with a single pressure pulse, while in our
case the interface is continuously driven. The use of a constant
amplitude $a_0$ in the rate equation (\ref{growthrm}) is a rough
approximation, as the amplitude of surface undulations grows
between each pressure oscillation. We therefore note $k a_{0}$ as
the free parameter in our model, and takes into account our
ignorance of the details of the solid-superfluid interface
dynamics. Note that the agreement is better for the case of
melting with an acoustic pulse from the liquid side (Fig.1).

Our model further predicts that $a_{RM}\propto \delta v_{m}
\propto \sqrt{I}$, while the radiation-pressure driven growth
(\ref{amppres}) predicts a linear dependence on the acoustic
power. In the regime where the melting is dominated by the
RM-instability, i.e. at relatively high temperatures, we therefore
expect to find $a_{RM}^2 \propto I$ behavior. The experimental
data at 1.2K \cite{nomura0} does indeed show such a power law
(Fig.3).

To conclude, we have analyzed recent experiments where the
solid-superfluid interface is accelerated by an acoustic-pulse,
and a large source of unexplained melting was observed. It was
suggested that some mechanism of energy dissipation at the
interface is responsible for this unexplained melting
\cite{nomura}. We proposed here a possible mechanism for such a
dissipation; local solid-liquid mixing (and melting) occurs due to
a dynamic instability of the interface, of the RM-type. The
proposed model is in good agreement with the experimental data,
and describes a specific mechanism by which the accelerated
interface melts the solid.

It will be interesting to study this instability using simulations
\cite{rychtmyer}, taking into account both the material flows and
the solid-liquid melting/solidification dynamics. Experiments on
faceted crystals, with an atomically smooth interface, should show
less melting due to the RM-like process described here. It could
also be useful to repeat these experiments with a controlled
initial undulation on the solid-liquid interface, of macroscopic
size, and visualize its evolution when accelerated by an acoustic
pulse. This initial undulation can be formed by a standing
capillary wave at the interface \cite{capilary}.

\begin{acknowledgments}
I wish to thank R. Nomura and Y. Okuda for access to their data,
many helpful discussions and their hospitality at the ULT
conference in Kanazawa, Japan (2002). This research is being
supported by the Louis and Anita Perlman Family Foundation and the
Koshland Postdoctoral Fellowship.
\end{acknowledgments}

\newpage

\begin{figure}\begin{center}
\centerline{\ \epsfysize 7cm \epsfbox{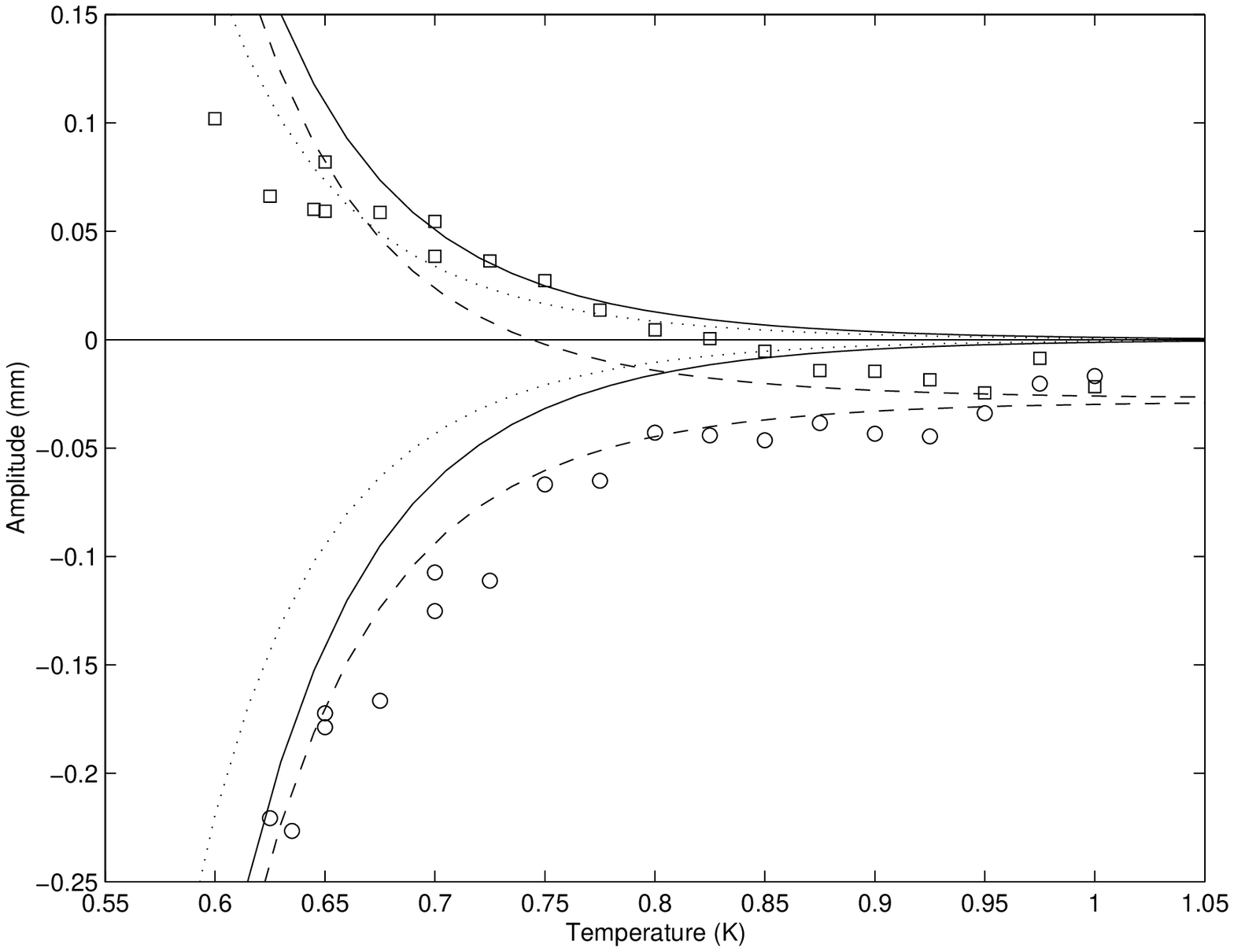}} \vskip 10mm
\caption{The measured melting/growth due to an acoustic pulse from
the solid/liquid side \cite{nomura} (circles \& squares
respectively). We plot the calculated radiation-pressure driven
growth/melt $h$ due to the acoustic pressure (Eq.\ref{amppres}),
using $\Delta=7.2$K (dotted lines) and $\Delta=7.5$K (solid lines)
in (Eq.\ref{growthcoef}). The dashed lines show the result of
subtracting the RM-amplitude $a_{RM}$ (since it represents extra
melting) (Eq.\ref{growthrm}) from the calculated linear
growth/melt ( Eq.\ref{growthcoef} using $\Delta=7.5$K, solid
lines), using $k a_{0}\simeq 20$.}
\end{center}
\end{figure}

\begin{figure}
\begin{center}
\centerline{\ \epsfysize 8cm \epsfbox{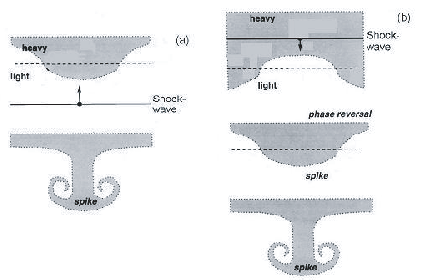}}
\caption{An illustration of the evolution of a Richtmyer-Meshkov
instability \cite{rychtmyerjpg} and the consequent material mixing
(bottom line). An initial sinusoidal undulation on the interface
(top line) is accelerated by an impacting shock-wave (horizontal
line with the attached arrow), from either the (a) light (low
density) or (b) heavy (high density) side. The inertial mixing
shown here ("spike" in the bottom line), after an acceleration by
a single pressure pulse, corresponds to the case of negligible
surface tension, unlike the solid-superfluid $^4$He case.}
\end{center}
\end{figure}

\begin{figure}
\centerline{\ \epsfysize 10cm \epsfbox{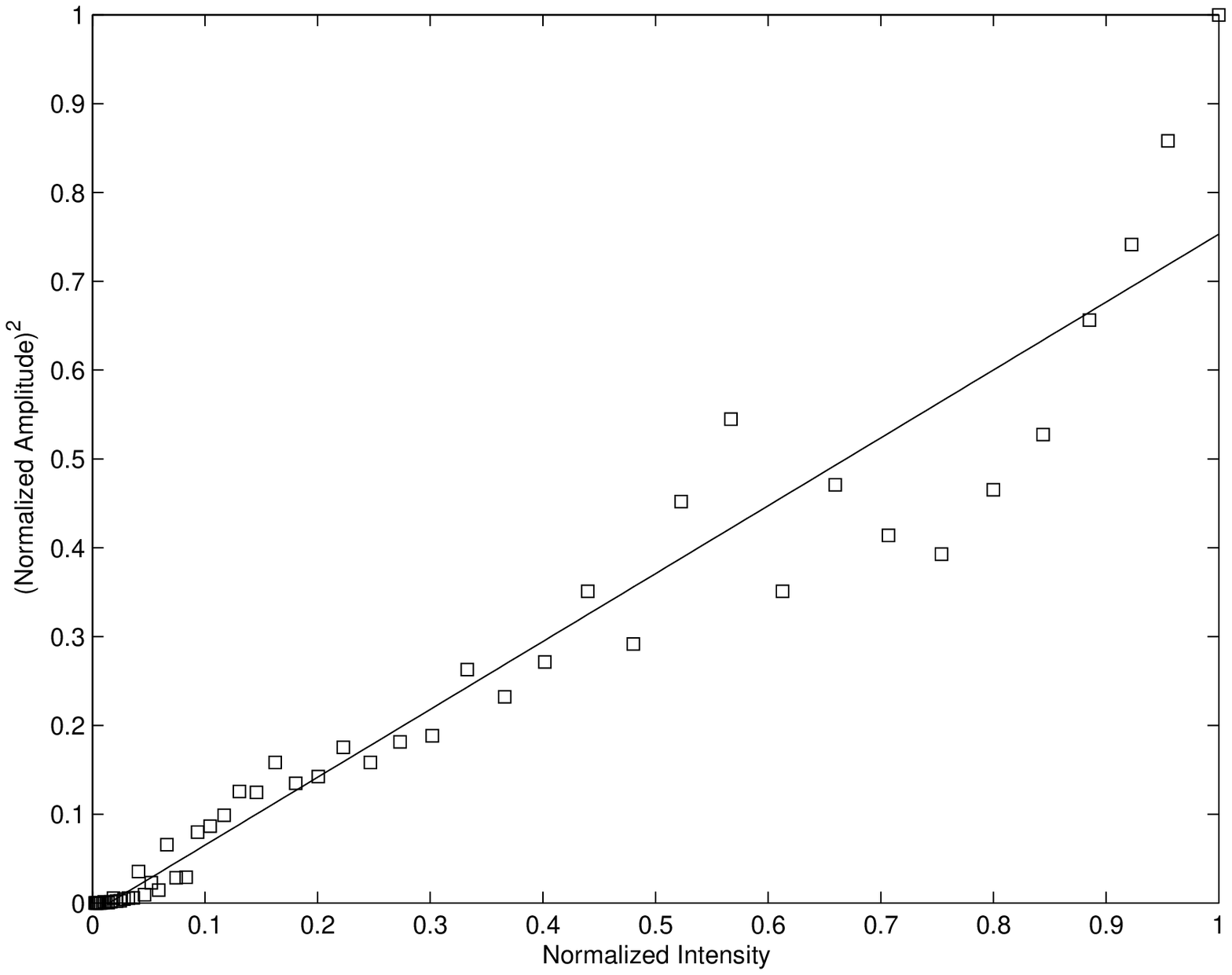}} \vskip 10mm
\caption{Measured melting depth as a function of the acoustic
power (both normalized by their largest measured values) at T=1.2K
(squares \cite{nomura0}). The straight line is a fit to illustrate
the $a_{RM}^2 \propto I$ behavior, as predicted by our model.}
\end{figure}


\begin{references}
\bibitem{note} Note that in order to get a good fit, using the
temperature-independent melting amplitude $a_{RM}$
(Eq.(\ref{growthrm})), we used $\Delta=7.5$K in
Eq.(\ref{growthcoef}), which is within the experimental
uncertainty \cite{bowley}.
\end{references}
\end{document}